 \documentstyle[12pt]{article}
\def\3{\ss}
\renewcommand{\theequation}{\thesection.\arabic{equation}}

\begin{document}
%
\makeatletter
\@addtoreset{equation}{section}
\makeatother
\renewcommand{\theequation}{\thesection.\arabic{equation}}
\begin{titlepage}
\thispagestyle{empty}

\begin{minipage}{13.25cm}

\vspace{-2cm}

\begin{flushright}
   HLRZ 92-34  \\
 FERMILAB-PUB-92/194-T \\
 BUTP-92/35  \\
   August, 1992
\end{flushright}

\begin{center}
   \begin{LARGE}
      {\bf{Continuum Gauge Fields from Lattice Gauge Fields}}
    \end{LARGE} \\[0.50cm]
    M. G\"ockeler$^{1,2}$, A. S. Kronfeld$^3$, G. Schierholz$^{2,4}$
    and U.-J. Wiese$^{5}$  \\[1.5em]
    {\small $^1$Institut f\"ur Theoretische Physik, RWTH Aachen,}
    \\[-0.25em]
    {\small Sommerfeldstra\3e, D-5100 Aachen, Germany}
    \\
    {\small $^2$Gruppe Theorie der Elementarteilchen,}
    \\[-0.25em]
    {\small H\"ochstleistungsrechenzentrum HLRZ,}
    \\[-0.25em]
    {\small c/o Forschungszentrum J\"ulich,
             D-5170 J\"ulich, Germany}
             \\
    {\small $^3$Theoretical Physics Group,
     Fermi National Accelerator Laboratory,}
    \\[-0.25em]
    {\small  P.O. Box 500, Batavia, Illinois 60510, U.S.A. }
    \\
    {\small $^4$Deutsches Elektronen-Synchrotron DESY,}
    \\[-0.25em]
    {\small Notkestra\3e 85, D-2000 Hamburg 52, Germany}
    \\
    {\small $^5$Institut f\"ur Theoretische Physik,
    Universit\"at Bern,}
    \\[-0.25em]
    {\small Sidlerstrasse 5, CH-3012 Bern, Switzerland}
\end{center}

\end{minipage}

\vspace{2cm}

\centerline{\bf Abstract}
\begin{quote}
On the lattice some of the salient features of pure
gauge theories and of
gauge theories with fermions in complex representations of the gauge
group seem to be lost. These features can be recovered by considering
part of the theory in the continuum. The prerequisite for that is
the construction of continuum gauge fields from lattice gauge fields.
Such a construction, which is gauge covariant
and complies with geometrical constructions
of the topological charge on the lattice, is given in this paper.
The procedure is explicitly carried out in the $U(1)$ theory in two
dimensions, where it leads to simple results.
\end{quote}

\end{titlepage}

\section{Introduction}

At present the most attractive nonperturbative regulator of
field theories is the lattice.
In this approach, matter fields are located on the lattice sites, and
gauge fields are represented by parallel transporters on the links
connecting the sites.
Usually it is most practical to work on a hypercubic lattice.

For many questions it is productive to envision and define properties of
an underlying continuum field.
The topological charge is a well-known example \cite{Lue}.
The geometrical object which determines the topological charge is a
principal fiber bundle.
To construct this bundle one needs to define transition functions
throughout the boundary of each elementary hypercube, i.e.\ in the
interior of each cube, using the lattice gauge field as input.
Under certain continuity assumptions on the lattice gauge field, the
bundle reconstructed is unique \cite{Lue,Phil}.
In other circumstances, one would like to define the gauge field itself
in the interior of the cubes and even the hypercubes.
Let us give a few examples:

As is well known, the lattice formulation of chiral gauge theories faces
great difficulties, which stem from the doubling problem \cite{Nielsen}.
To overcome this problem we have proposed a method \cite{Goe} which
keeps the fermionic degrees of freedom in the continuum.\footnote{For
similar ideas see also ref.~\cite{Jan}.}
The motivation comes from considering the interplay between chiral
symmetries, anomalies, and the doubling problem.
Anomalies may arise when the ultraviolet regulator breaks some of the
action's symmetries.
For vector-like lattice theories chiral symmetry-breaking terms in the
lattice action solve the doubling problem, {\em and\/} produce the
abelian chiral anomaly \cite{Karsten}.
(As for the nonabelian anomaly, see ref.~\cite{Coste}.)
Without these terms, e.g.\ for supposed-to-be chiral fermions,
the doublers cancel the
anomalies in Noether currents, to which gauge fields can be coupled.
Usually anomaly cancellation is a desired feature, but here it is
unacceptable for several reasons.
First, the pattern is wrong, because the cancellation is among doublers
and not among different fermion species.
Second, physically desirable anomalies in global symmetries
might also be eliminated.
Finally, the doublers behave like a multiplet of particles with an
equal number of right- and left-handed low-energy fermions;
they make the lattice theory vector-like.

In the continuum path integral formalism, anomalous symmetries are
broken by the fermion measure \cite{Fuji}.
This formalism is a more flexible approach.
Gauged symmetries can be rendered non-anomalous either through anomaly
cancellation among different flavors (as in the standard model) or by
choosing anomaly free representations (as in GUT's).
When the measure is correctly defined \cite{A-GG}, it is also
possible to obtain consistent anomalies for global symmetries.
In ref.~\cite{Goe} we suggest treating fermions in this way, even when
the gauge fields are regulated by the lattice.\footnote{The price is a
separate ultraviolet regulator for the fermions.}
The most transparent way to proceed is to construct continuum gauge
fields from lattice gauge fields.

In vector theories, a related problem arises in the context of the
Atiyah-Singer index theorem~\cite{Atiyah}.
This theorem, which holds for smooth gauge fields, says that the number
of right-handed minus left-handed zero modes of the Dirac operator
equals the topological charge of the gauge field, thus connecting the
topology of gauge fields with the chiral properties of the fermions.
On the lattice the index theorem is lost whenever the discretized Dirac
operator has no geometric meaning.
For example, both the staggered and Wilson formulations have
no exact zero modes and one must be satisfied with a remnant of the
index theorem \cite{Smit}.
On the other hand, the index theorem holds for continuum fermions in the
continuum gauge fields constructed in this paper.
The zero modes could be studied numerically by re-introducing finer and
finer sublattices.

Another observable, which has become of interest recently in the context
of baryon number violating processes \cite{Bar} in the standard
model, is the Chern-Simons number \cite{Cher}. One of its characteristic
features is that it changes by an integer
under ``large'' gauge transformations, the integer being the winding
number of the gauge transformation.
A lattice construction of the Chern-Simons
number should share this feature of the continuum
expression. This is not the case for the naively discretized expression,
so that a more refined treatment is necessary. Again, knowing the
continuum gauge fields, we may define the Chern-Simons number
exactly as in the continuum \footnote{For different constructions see,
however, refs.~\cite{Sei,Woi,PS90}.}.

In this paper we shall construct a continuum gauge field, $A_\mu$,
from a lattice gauge field.
Our continuum gauge field fulfills the following constraints: (i) The
parallel transporters derived from $A_\mu$
agree with those of the original lattice gauge field.
(ii) A lattice gauge transformation results in a continuum
gauge transformation of $A_\mu$, i.e.\ the construction of $A_\mu$
is gauge covariant. (iii) The gauge field $A_\mu$
is a connection in the fiber bundle \cite{Lue}
that has been constructed from the
lattice gauge field.
We consider these three criteria essential for consistency.
In particular, the third feature guarantees that one obtains the same
value for the topological charge, independent of whether one computes it
from the transition functions or by integrating the Chern-Pontryagin
density.

The paper is organized as follows. In sec.~2 we derive the continuum
gauge field in four dimensions. Some extensions of older results,
which we need here, are described in Appendix A.
We then apply our results
to the $U(1)$ gauge theory in two dimensions in sec.~3. Finally, in
sec.~4 we conclude with some remarks.

\section{Continuum gauge field}

We now turn to the construction of the continuum gauge field
$A_\mu = - A^+_\mu$.
For definiteness we consider the gauge group $SU(N)$.
We shall restrict ourselves to a hypercubic, periodic lattice
of period $L$, i.e.\
the underlying continuum manifold is the four-torus ${\bf T}^4$. The
lattice is then defined by
\begin{equation}
\Lambda = \{ s \in {\bf T}^4 \mid s_\mu \in {\bf Z}, \forall \mu \}.
\end{equation}
We denote the parallel transporters by $U(s,\mu)$.

The construction proceeds in two steps. In the first step
we construct in each hypercube,
\begin{equation}
c(s) = \{ x \in {\bf T}^4 \mid s_\mu \leq x_\mu \leq s_\mu + 1,
\forall \mu \},
\end{equation}
a vector field $A_\mu^{(s)}$ such that on the intersection of two
adjacent hypercubes the corresponding fields are connected by a
gauge transformation given by L\"uscher's~\cite{Lue} transition
functions. In the second step all these gauge fields will
be brought into the same gauge such that one (possibly singular)
gauge field on the whole torus emerges.

Since the gauge transformation of a field involves
derivatives, it turns out to be necessary to enlarge the hypercubes
$c(s)$ to open sets
\begin{equation}
\tilde{c}(s) := \{ x \in {\bf T}^4 \mid - \varepsilon < x_\mu - s_\mu
< 1 + \varepsilon, \forall \mu \}
\end{equation}
with $0 < \varepsilon < 1/2$. The intersection of two neighboring
hypercubes is denoted by
\begin{equation}
\tilde{f}(s,\mu) := \tilde{c}(s) \cap \tilde{c}(s-\hat{\mu})
= \{ x \in \tilde{c}(s) \mid - \varepsilon < x_\mu - s_\mu
< \varepsilon \}.
\end{equation}
Furthermore, we define for $\mu \neq \nu$
\begin{eqnarray}
\tilde{p}(s,\mu,\nu) &:=& \tilde{c}(s) \cap \tilde{c}(s - \hat{\mu})
\cap \tilde{c}(s - \hat{\nu}) \cap \tilde{c}(s - \hat{\mu} - \hat{\nu})
\nonumber \\
&=& \{ x \in \tilde{c}(s) \mid - \varepsilon < x_\mu - s_\mu
< \varepsilon, - \varepsilon < x_\nu - s_\nu < \varepsilon \}.
\end{eqnarray}
{}From
L\"uscher's transition functions
\begin{equation}
v_{s,\mu} : f(s,\mu) \rightarrow SU(N),
\end{equation}
defined on the faces $f(s,\mu) = c(s) \cap c(s - \hat{\mu})$, we
construct ``smeared'' transition functions
\begin{equation}
\tilde{v}_{s,\mu}: \tilde{f}(s,\mu) \rightarrow SU(N).
\end{equation}
Let $\phi: {\bf R} \rightarrow [0,1]$ be smooth and such that
\begin{eqnarray}
\phi(t) & = 0 & {\rm if}\: t < \eta,\\
\phi(t) & = 1 & {\rm if}\: t > 1 - \eta,
\end{eqnarray}
where $\varepsilon < \eta < 1/2$. We then define
\begin{equation}
\tilde{v}_{s,\mu}(x) = v_{s,\mu}(s_1 + \phi(x_1 -
s_1),\cdots,s_\mu,\cdots,s_4 + \phi(x_4 - s_4))
\label{tran}
\end{equation}
(Here and in the following $f(\cdots, s_\alpha,\cdots)$ means that the
function $f$ is to be evaluated with $x_\alpha$ put equal to $s_\alpha$,
irrespective of the order in which the arguments are written.) From the
fact that L\"uscher's transition functions satisfy the cocycle
condition
\begin{equation}
v_{s - \hat{\mu},\nu}(x) v_{s,\mu}(x)
= v_{s - \hat{\nu},\mu}(x) v_{s,\nu}(x)
\end{equation}
on
$p(s,\mu,\nu) = c(s) \cap c(s - \hat{\mu})
\cap c(s - \hat{\nu}) \cap c(s - \hat{\mu} - \hat{\nu})$
it follows that on $\tilde{p}(s,\mu,\nu)$ the cocycle condition for
$\tilde{v}_{s,\mu}$ is fulfilled. Therefore the transition functions
$\tilde{v}$ define an $SU(N)$ principal bundle over ${\bf T}^4$.

Now we introduce the abbreviations
\begin{eqnarray}
Z_\lambda(s,\alpha,\beta,\gamma \mid x) & := &
(\tilde{v}_{s+\hat{\gamma},\gamma} \tilde{v}_{s+\hat{\gamma}+
\hat{\beta},\beta} \tilde{v}_{s+\hat{\gamma}+\hat{\beta}+\hat{\alpha},
\alpha} \nonumber \\
&\times& \partial_\lambda \tilde{v}^{-1}_{s+\hat{\gamma}+\hat{\beta}
+\hat{\alpha},\alpha} \tilde{v}^{-1}_{s+\hat{\gamma}+\hat{\beta},\beta}
\tilde{v}^{-1}_{s+\hat{\gamma},\gamma}) (x),\\
Z_\lambda(s,\alpha,\beta \mid x) & := &
(\tilde{v}_{s+\hat{\alpha},\alpha} \tilde{v}_{s+\hat{\alpha}+
\hat{\beta},\beta}
\partial_\lambda \tilde{v}^{-1}_{s+\hat{\alpha}+\hat{\beta},
\beta} \tilde{v}^{-1}_{s+\hat{\alpha},\alpha}) (x),\\
Z_\lambda(s,\alpha \mid x) & := & (\tilde{v}_{s+\hat{\alpha},\alpha}
\partial_\lambda \tilde{v}^{-1}_{s+\hat{\alpha},\alpha}) (x),\\
{\rm Ad}(\tilde{v}) M & := & \tilde{v} M \tilde{v}^{-1}.
\end{eqnarray}
Due to the cocycle condition, $Z_\lambda(s,\alpha,\beta,\gamma \mid x)$
and
$Z_\lambda(s,\alpha,\beta \mid x)$ are symmetric in $\alpha, \beta,
\gamma$ and $\alpha, \beta$, respectively.
Furthermore, one has relations of the type
\begin{equation}
Z_\lambda(s,\alpha,\beta,\gamma \mid x) = {\rm Ad}(\tilde{v}_{s+\hat{
\gamma},\gamma}(x)) Z_\lambda(s+\hat{\gamma},\alpha,\beta \mid x)
+ Z_\lambda(s,\gamma \mid x).
\end{equation}
For $x \in \tilde{c}(s)$ and $\alpha, \beta, \gamma$ the indices
complementary to $\mu$ we define the field $A_\mu^{(s)}$ by
\begin{equation} \begin{array}{l} \displaystyle
A_\mu^{(s)}(x) = \phi(x_\alpha - s_\alpha) \phi(x_\beta -s_\beta)
\phi(x_\gamma - s_\gamma)
      \{ Z_\mu(s,\alpha,\beta,\gamma \mid s_\alpha + 1,
s_\beta + 1,s_\gamma + 1,x_\mu)
\\   \displaystyle
                + \sum_{\stackrel{\rm cycl.\, perm.}{(\alpha,\beta,
\gamma)}}       [ -{\rm Ad}     ((\tilde{v}_{s+\hat{\alpha},\alpha}
\tilde{v}_{s+\hat{\alpha}+\hat{\beta},\beta})(s_\alpha + 1,s_\beta
+ 1,x_\gamma,x_\mu)       )
\\   \displaystyle
\hphantom{   + \sum_{\stackrel{\rm cycl.\, perm.}{(\alpha,\beta,
\gamma)}} } \qquad
\times Z_\mu(s+\hat{\alpha}+\hat{\beta},\gamma \mid
s_\alpha + 1,s_\beta + 1,s_\gamma + 1,x_\mu)
\\   \displaystyle
\hphantom{   + \sum_{\stackrel{\rm cycl.\, perm.}{(\alpha,\beta,
\gamma)}} }
+{\rm Ad}      (\tilde{v}_{s+\hat{\alpha},\alpha}(s_\alpha + 1,
x_\beta, x_\gamma, x_\mu)
\tilde{v}_{s+\hat{\alpha}+\hat{\beta},\beta}(s_\alpha + 1,s_\beta
+ 1,x_\gamma,x_\mu)       )
\\   \displaystyle
\hphantom{   + \sum_{\stackrel{\rm cycl.\, perm.}{(\alpha,\beta,
\gamma)}} } \qquad
\times Z_\mu(s+\hat{\alpha}+\hat{\beta},\gamma \mid
s_\alpha + 1,s_\beta + 1,s_\gamma + 1,x_\mu)
\\   \displaystyle
\hphantom{   + \sum_{\stackrel{\rm cycl.\, perm.}{(\alpha,\beta,
\gamma)}} }
+{\rm Ad}      (\tilde{v}_{s+\hat{\alpha},\alpha}(s_\alpha + 1,
x_\beta, x_\gamma, x_\mu)
\tilde{v}_{s+\hat{\alpha}+\hat{\gamma},\gamma}(s_\alpha + 1,x_\beta,
s_\gamma + 1,x_\mu)       )
\\   \displaystyle
\hphantom{   + \sum_{\stackrel{\rm cycl.\, perm.}{(\alpha,\beta,
\gamma)}} }  \qquad
\times Z_\mu(s+\hat{\alpha}+\hat{\gamma},\beta \mid
s_\alpha + 1,s_\beta + 1,s_\gamma + 1,x_\mu)
\\   \displaystyle
\hphantom{   + \sum_{\stackrel{\rm cycl.\, perm.}{(\alpha,\beta,
\gamma)}} }
-{\rm Ad}      (\tilde{v}_{s+\hat{\alpha},\alpha}(s_\alpha + 1,
x_\beta, x_\gamma, x_\mu)       )
\\   \displaystyle
\hphantom{   + \sum_{\stackrel{\rm cycl.\, perm.}{(\alpha,\beta,
\gamma)}} }  \qquad
\times Z_\mu(s+\hat{\alpha},\beta,\gamma \mid
s_\alpha + 1,s_\beta + 1,s_\gamma + 1,x_\mu)       ]       \}
\\   \displaystyle
 + \sum_{\stackrel{\rm cycl.\, perm.}{(\alpha,\beta,\gamma)}}
      [ - \phi(x_\alpha - s_\alpha) \phi(x_\beta -s_\beta)
Z_\mu(s,\alpha,\beta \mid
s_\alpha + 1,s_\beta + 1,x_\gamma ,x_\mu)
\\   \displaystyle
\hphantom{   + \sum_{\stackrel{\rm cycl.\, perm.}{(\alpha,\beta,
\gamma)}} }
+ \phi(x_\alpha - s_\alpha) \phi(x_\beta -s_\beta)
{\rm Ad}      (\tilde{v}_{s+\hat{\alpha},\alpha}(s_\alpha + 1,
x_\beta, x_\gamma, x_\mu)       )
\\   \displaystyle
\hphantom{   + \sum_{\stackrel{\rm cycl.\, perm.}{(\alpha,\beta,
\gamma)}} } \qquad
\times Z_\mu(s+\hat{\alpha},\beta \mid
s_\alpha + 1,s_\beta + 1,x_\gamma ,x_\mu)
\\   \displaystyle
\hphantom{   + \sum_{\stackrel{\rm cycl.\, perm.}{(\alpha,\beta,
\gamma)}} }
+ \phi(x_\alpha - s_\alpha) \phi(x_\gamma -s_\gamma)
{\rm Ad}      (\tilde{v}_{s+\hat{\alpha},\alpha}(s_\alpha + 1,
x_\beta, x_\gamma, x_\mu)       )
\\   \displaystyle
\hphantom{   + \sum_{\stackrel{\rm cycl.\, perm.}{(\alpha,\beta,
\gamma)}} }  \qquad
\times Z_\mu(s+\hat{\alpha},\gamma \mid
s_\alpha + 1,x_\beta ,s_\gamma + 1,x_\mu)
\\   \displaystyle
\hphantom{   + \sum_{\stackrel{\rm cycl.\, perm.}{(\alpha,\beta,
\gamma)}} }
+ \phi(x_\alpha - s_\alpha)
Z_\mu(s,\alpha \mid
s_\alpha + 1,x_\beta,x_\gamma ,x_\mu)       ]  \,.
\end{array}
\label{pot}
\end{equation}
It is straightforward to verify that this field has the desired
properties:
\begin{equation}
A_\lambda^{(s-\hat{\mu})}(x) = \tilde{v}_{s,\mu}(x) A_\lambda^{(s)}(x)
\tilde{v}^{-1}_{s,\mu}(x) + \tilde{v}_{s,\mu}(x) \partial_\lambda
\tilde{v}^{-1}_{s,\mu}(x)
\label{patch}
\end{equation}
for $x \in \tilde{f}(s,\mu)$ and
\begin{equation}
{\cal P}\, {\rm exp}
\left \{ \int_0^1 dt A_\mu^{(s)}(x+(1-t)\hat{\mu}) \right \}
= u^s_{x,x+\hat{\mu}},
\end{equation}
where $x$ and $x+\hat{\mu}$ are corners of $c(s)$ and $u^s_{x,x+\hat
{\mu}}$ are the link variables in the complete axial gauge defined
in ref.~\cite{Lue}.

The field (\ref{pot})
can be constructed step by step starting from the value zero on the
neighborhood $\{x \in \tilde{c}(s) \mid -\varepsilon < x_\lambda
- s_\lambda < \varepsilon, \forall \lambda\}$ of the lattice point $s$.
Equation (\ref{patch}) then gives the field on the analogous
neighborhoods of the other corners of $c(s)$. Since according to
eq.\ (\ref{tran}) $\tilde{v}$ is constant on these neighborhoods, one
gets again zero. In the next step one puts the field equal to zero
on $\varepsilon$-neighborhoods of the links in $c(s)$ originating
from $s$ and applies (\ref{patch}) again. Then one defines
$A_\lambda^{(s)}$ on the six $\tilde{p}(s,\mu,\nu)$ by interpolating
the expressions already obtained. By means of (\ref{patch}) one finds
$A_\lambda^{(s)}$ on all the $\tilde{p}$'s contained in $c(s)$.
Continuing in this fashion one arrives at eq.\ (\ref{pot}).

It should be remarked that our construction works only if the
transition functions $v_{s,\mu}$ are well defined. This is not the
case for the exceptional configurations as defined by L\"uscher.
However, since these form only a set of measure zero in the
functional integral, they can be ignored. Furthermore, covariance
under hypercubic rotations and reflections is already violated by
the construction of the transition functions, hence also by our
expression for the continuum gauge field $A^{(s)}_\mu$.

The purpose of $\varepsilon$ and the function $\phi$ is only to give
a well defined meaning to $v_{s,\mu}\partial_\lambda v^{-1}_{s,\mu}$
even for
$\lambda = \mu$ in the course of the construction. In the final
expression (\ref{pot}) for $A_\mu^{(s)}$ one can
perform the limit $\varepsilon, \eta \rightarrow 0$ and put $\phi(t)=t$
for $0 \leq t \leq 1$ without problems.

We now come to the second step, i.e.\ we construct a ``global''
(but possibly singular) field $A_\mu$.
To do so one can make use of the section $w^s$ constructed
in ref.~\cite{int} on the boundaries of the hypercubes.
In Appendix A we extend $w^s$ into the interior of $c(s)$.
In this procedure one will, in general, encounter singularities.
Nevertheless, one can define on $c(s)$
\begin{equation}
A_\mu(x) = w^s(x)^{-1} A_\mu^{(s)}(x) w^s(x) + w^s(x)^{-1} \partial_\mu
w^s(x).
\label{glopo}
\end{equation}
Under a lattice gauge transformation
\begin{equation}
U(s,\mu) \to
\bar{U}(s,\mu) = g(s) U(s,\mu) g(s+\hat{\mu})^{-1}
\end{equation}
the gauge field $A_\mu^{(s)}$ transforms as
\begin{equation}
\bar{A}_\mu^{(s)}(x) = g(s) A_\mu^{(s)}(x) g(s)^{-1} \,,
\end{equation}
and the section $w^s$ transforms as
\begin{equation}
\bar{w}^s(x) = g(s) w^s(x) g(x)^{-1},
\end{equation}
where $g(x)$ is the interpolation of $\{ g(s) \}$ to all of $c(s)$
which we give in appendix A.   This entails the desired behavior
\begin{equation}
\bar{A}_\mu(x) = g(x) (A_\mu(x) + \partial_\mu) g(x)^{-1}.
\end{equation}

\section{$U(1)$ gauge theory in two dimensions}

Since our construction of the continuum gauge field is rather
involved, it may be useful to repeat the procedure for a $U(1)$-theory
in two dimensions, where it is much simpler. In particular, it will
be possible to discuss the ``global'' field (\ref{glopo})
explicitly.

Using the obvious analogues of the concepts introduced in sec.\ 2
for the case of four dimensions we get for L\"uscher's transition
functions
\begin{eqnarray}
v_{s,1} (x) & = & U(s- \hat{1},1) \quad,\quad x \in f(s,1) \,,
\nonumber \\
v_{s,2} (x) & = & U(s- \hat{2},2)  P(s-\hat{2})
^{x_1 - s_1} \quad,\quad x \in f(s,2) \,.
\end{eqnarray}
Here $P(s)$ denotes the plaquette variable
\begin{equation}
P(s) = U(s,1) U(s+\hat{1},2) U(s+\hat{2},1)^{-1} U(s,2)^{-1} \,.
\end{equation}
The noninteger power of a group element $e^{i \alpha}$ is defined by
\begin{equation}
     ( e^{i \alpha}       ) ^t = e^{it \alpha}\,,
\end{equation}
where $0 \leq t \leq 1$ and $\alpha$ is restricted to the interval
$- \pi < \alpha < \pi$. If $e^{i \alpha} = -1$, the power
$\left( e^{i \alpha} \right)^t$ is not  well-defined. Configurations
with at least one plaquette variable equal to $-1$ are exceptional
in the sense of L\"uscher and do not allow our constructions to
be performed.

Using the same smearing function $\phi$ as in four dimensions one
obtains the transition functions
\begin{equation}
\tilde{v}_{s,\mu}: \tilde{f}(s,\mu) \rightarrow U(1)\,.
\end{equation}
The step-by-step   construction indicated in sec.\ 2 leads to
the field
\begin{eqnarray}
A_1 ^{(s)} (x) &=& i \phi^\prime (x_1-s_1) \phi(x_2-s_2)
\log P(s)  \,, \nonumber \\
A_2 ^{(s)} (x) &=& 0
\end{eqnarray}
for $x \in \tilde{c} (s)$, where in accordance with our definition
of powers we have defined
\begin{equation}
\log      (e^{i \alpha}       ) = i \alpha \quad,\quad
- \pi < \alpha < \pi\,.
\end{equation}
Note that the field has been chosen real as is commonly done
for gauge group $U(1)$.

It is now easy to check that
\begin{equation}
A_\lambda^{(s-\hat{\mu})}(x) =  A_\lambda^{(s)}(x)
+ i \tilde{v}^{-1}_{s,\mu}(x) \partial_\lambda
\tilde{v}_{s,\mu}(x)
\end{equation}
for $x \in \tilde{f} (s,\mu)$.
For the field strength in $\tilde{c}(s)$ one finds
\begin{equation}
F_{12} (x) = -i \phi^\prime(x_1-s_1) \phi^\prime(s_2-s_2)
\log P(s)
\end{equation}
so that in the limit $\eta \to 0$, $\phi (t) = t$, one ends up with
a constant field strength in each plaquette as was already considered
by Flume and Wyler \cite{fluwy}. In this limit the field is
given by
\begin{eqnarray}
A_1 ^{(s)} (x) &=& i (x_2 - s_2) \log P(s)  \,,\nonumber \\
A_2 ^{(s)} (x) &=& 0 \,.
\end{eqnarray}

In order to obtain a ``global'' field we have to construct the
section $w^s$ and to discuss its singularities in the interior of
$c(s)$. Straightforward adaptation of the appropriate formulas of
ref.~\cite{int} to the case of two dimensions yields on the
boundary of $c(s)$:
\begin{eqnarray}
w^s (s_1,x_2) &=& U(s,2)^{x_2-s_2} \,, \nonumber \\
w^s (s_1+1,x_2) &=& U(s+\hat{1},2)^{x_2-s_2} U(s,1)\,, \nonumber \\
w^s (x_1,s_2) &=& U(s,1)^{x_1-s_1} \,, \nonumber \\
w^s (x_1,s_2+1) &=& P(s)^{x_1-s_1} U(s+\hat{2},1)^{x_1-s_1} U(s,2)\,.
\end{eqnarray}
Again, certain configurations (those with at least one link
variable equal to $-1$) have to be excluded. But this should cause
no harm because they form a set of measure zero in the functional
integral. The interpolation into the interior of $c(s)$ is
however more subtle, since one will necessarily encounter
singularities if the winding number of the map
\begin{equation}
w^s |_{\partial c(s)} : \partial c(s) \to U(1)
\label{map}
\end{equation}
is nonzero. The corresponding configurations can no longer be
ignored because, for example, all configurations with nonvanishing
topological charge belong to this class. Nevertheless, proceeding
as outlined in appendix A one obtains
\begin{equation} \begin{array}{l} \displaystyle
w^s (x_1,x_2) = w^s (s_1,x_2) [ w^s (s_1,x_2) ^{-1}
w^s (s_1+1,x_2) (P(s)^{-1} )^{x_2-s_2} ]
^{x_1-s_1} \\ \displaystyle
\phantom{w^s (x_1,x_2) = w^s (s_1,x_2)}
\times [ P(s)^{x_2-s_2} ] ^{x_1-s_1} \,.
\end{array} \end{equation}
This expression is  ill-defined whenever
\begin{equation}
w^s (s_1,x_2) ^{-1}
w^s (s_1+1,x_2)      (P(s)^{-1}       )^{x_2-s_2} = -1 \,.
\end{equation}

To make the construction more transparent we write
\begin{equation} \begin{array}{rcll} \displaystyle
U(x,\mu) & = & e^{i \psi(x,\mu)}
& \quad ,\, - \pi < \psi (x,\mu) < \pi \,, \\ \displaystyle
P(s) & = & e^{i \omega (s) }
& \quad ,\, - \pi < \omega(s) < \pi \,.
\end{array}
\end{equation}
Furthermore, we introduce the function $q(\alpha)$ which shifts an
angle into the interval $-\pi < q(\alpha) < \pi$ by adding integer
multiples of $2 \pi$:
\begin{equation}
q(\alpha) = \sum_{n=- \infty} ^{\infty} (\alpha - 2 \pi n)
\theta (\alpha - 2 \pi n + \pi) \theta(\pi - \alpha + 2 \pi n) \,.
\end{equation}
So we have
\begin{equation}
\omega (s) = q      ( \psi(s,1) + \psi(s+\hat{1},2) -
\psi(s+\hat{2},1) - \psi(s,2)       )
\end{equation}
and
\begin{equation} \begin{array}{l} \displaystyle
w^s (x_1,x_2) = e^{i \psi(s,2) (x_2-s_2)}
e^{i \omega(s) (x_1-s_1) (x_2-s_2)}
\\ \displaystyle  \phantom{ w^s (x_1,x_2) = } \times
e^{i (x_1-s_1) q \left( (-\psi(s,2) + \psi(s+\hat{1},2) - \omega(s))
(x_2-s_2) + \psi(s,1) \right) } \,.
\end{array} \end{equation}
The last factor will jump from $e^{i \pi (x_1 - s_1)}$ to
$e^{-i \pi (x_1 - s_1)}$ (or the other way around) if $x_2$
passes through a value with
\begin{equation}
(-\psi(s,2) + \psi(s+\hat{1},2) - \omega(s))
(x_2-s_2) + \psi(s,1) = (2n+1) \pi \quad,
\quad n \,\,\, \mbox{integer} \,.
\end{equation}
Away from these gauge singularities the field
\begin{equation}
A_\mu(x) = A_\mu^{(s)}(x) -i w^s(x)^{-1} \partial_\mu
w^s(x) \quad , \, x \in c(s) \,,
\end{equation}
can be written as
\begin{eqnarray}
A_1 (x) &=& q      ( \psi(s,1) + (- \psi(s,2) + \psi(s+\hat{1},2)
- \omega(s)) (x_2-s_2)       ) \,, \nonumber \\
A_2 (x) &=& \psi(s,2) + (- \psi(s,2) + \psi(s+\hat{1},2))
(x_1-s_1) \,.
\label{pot2}
\end{eqnarray}
If the winding number of the map (\ref{map}) vanishes one has
\begin{equation}
\omega (s) = \psi(s,1) + \psi(s+\hat{1},2) - \psi(s+\hat{2},1)
- \psi(s,2) \,,
\end{equation}
and consequently
\begin{eqnarray}
A_1 (x) &=& \psi(s,1) + (- \psi(s,1) + \psi(s+\hat{2},1))
(x_2-s_2) \,, \nonumber \\
A_2 (x) &=& \psi(s,2) + (- \psi(s,2) + \psi(s+\hat{1},2))
(x_1-s_1) \,.
\end{eqnarray}
i.e.\ in this case the construction of $A_\mu (x)$ is symmetric with
respect to the interchange of the directions 1 and 2,
although in general the interpolation violates invariance under
$90^\circ$ rotations.

It is easy to see that integration of the field (\ref{pot2}) over the
links reproduces the link angles of the lattice gauge field, e.g.
\begin{equation}
\int_{s_1}^{s_1+1} \! dx_1 A_1 (x_1,s_2) = \psi (s,1) \,.
\end{equation}
One should however note that eq.(\ref{pot2})
gives $A_\mu (x)$ only within
$c(s)$. How do the expressions on neighboring plaquettes fit
together? One finds that $A_2 (x)$ is continuous as a function
of $x_1$ and $A_1 (x)$ is continuous as a function of $x_2$.
So one gets continuity in the transverse direction whereas in
the longitudinal direction the field will in general
exhibit jumps. But these can be gauged away and should not
cause any trouble.

Let us finally discuss the Chern-Simons number resulting from our
continuum gauge field.
In the continuum, the
Chern-Simons number is given by the line integral
\begin{equation}
n_{CS} = \frac{1}{2 \pi} \int \! dx_\mu A_\mu (x) \,.
\end{equation}
So using our interpolated field (\ref{pot2})
we find as the contribution
of one link the integral of $A_\mu (x)$ over this link (divided
by $2 \pi$). As remarked above, this integral reproduces the
original link angle:
\begin{equation}
\frac{1}{2 \pi} \int_s^{s+\hat{\mu}} \! dx_\mu A_\mu (x) =
\frac{1}{2 \pi} \psi (s,\mu) \,.
\end{equation}

 On the other hand, starting analogously to Seiberg \cite{Sei}
 from the section
 $w^s (x)$ on $\partial c (s)$ one arrives at the following
 expressions:
\begin{eqnarray}
\frac{-i}{2 \pi} \int_{s_1}^{s_1+1} \! dx_1 w^s (x) ^{-1}
\partial_1 w^s (x) |_{x_2=s_2} & = & \frac{1}{2 \pi} \psi(s,1) \,, \\
\frac{-i}{2 \pi} \int_{s_1}^{s_1+1} \! dx_1 w^s (x) ^{-1}
\partial_1 w^s (x) |_{x_2=s_2+1} & = &
\frac{1}{2 \pi}      ( \omega(s) + \psi(s+\hat{2},1)       ) \,,  \\
\frac{-i}{2 \pi} \int_{s_2}^{s_2+1} \! dx_2 w^s (x) ^{-1}
\partial_2 w^s (x) |_{x_1=s_1} & = & \frac{1}{2 \pi} \psi(s,2) \,, \\
\frac{-i}{2 \pi} \int_{s_2}^{s_2+1} \! dx_2 w^s (x) ^{-1}
\partial_2 w^s (x) |_{x_1=s_1+1} & = &
\frac{1}{2 \pi} \psi(s+\hat{1},2) \,.
\end{eqnarray}
 Hence one obtains immediately agreement between the two approaches
 on the links starting from the origin $s$ of the plaquette $c(s)$,
 whereas on the other links one gets coinciding results in general
 only in the naive continuum limit.
 (Re-introduce the lattice spacing $a$ and recall that the plaquette
 angle $\omega (s)$ should vanish as $a^2$, whereas the link angles
 $\psi (s,\mu)$ vanish as $a$ in the continuum limit.)

\section{Conclusions}

The great advantage of lattice gauge theory is that it provides a
framework for explicit nonperturbative calculations.
Unfortunately, it is not always easy to understand how other
nonperturbative information, derived by geometric or topological
methods, fits in with the lattice.
Our construction of a continuum gauge field from a lattice gauge field
should help bridge this gap.
Although the expressions appear rather complicated, we have obtained a
relatively simple result in the case of the two-dimensional U(1) theory.
In the future we hope to apply the four-dimensional results to chiral
fermions, as suggested in ref.~\cite{Goe}.

A crucial question concerns the influence of short-distance fluctuations
on the interpolation and on the quantities constructed from the
continuum gauge field.
{}From the example of the topological susceptibility it is known that such
fluctuations, the so-called dislocations \cite{dis}, can make
nonperturbative, ultraviolet divergent contributions to physical
quantities.
In this case the divergence can be controlled by choosing an appropriate
gauge field action.
It will be important to understand whether analogous problems arise with
quantities constructed out of the continuum gauge field, and, if so, how
to cope with them.

Having defined a continuum gauge field for each lattice gauge field, one
may use the continuum action of this field as the lattice gauge action.
Unless there is a construction in which the integrals can be carried out
analytically, the use of this action in a numerical simulation is
unpractical, but it has some theoretical advantages.
For example, one has the usual bound on the action in each topological
sector.
Then there is no nonperturbative divergence in the topological
susceptibility and, perhaps, in other quantities of interest.

In closing we mention some related results that we have obtained, but
that we do not present in this paper.
Our present discussion is based on L\"uscher's transition functions
\cite{Lue}, but we have also carried out the construction for the
Phillips-Stone bundle \cite{Phil}, which is defined on simplicial
lattices.
Furthermore, although this paper focuses on the gauge field, we
have obtained corresponding extrapolations for matter fields too.

\section*{Acknowledgements}
This work was supported in part by the
Deutsche Forschungsgemeinschaft and by the
Schweizerischer Nationalfonds.
The paper was completed while one of us (G. S.) was visiting Brookhaven
National Laboratory. He wants to thank the theory group for its warm
hospitality.

\appendix
\section{Interpolation of the section to the interior of the hypercube}

Here we describe how the section $w^s$ is interpolated from the
boundary of the hypercube $c(s)$ to its interior. The interpolation of
$w^s$ from the corners of the hypercube to its boundary was already
given in ref.~\cite{int}. We also investigate
the properties of the
section under lattice gauge transformations.

On the corners of the hypercube $c(s)$ the section is given by
\begin{equation}
w^s (x) = U(s,1)^{y_1} U(s+y_1 \hat{1},2)^{y_2}
 U(s+y_1 \hat{1} +y_2 \hat{2},3)^{y_3}
 U(s+y_1 \hat{1} +y_2 \hat{2} +y_3 \hat{3},4)^{y_4} \,,
\end{equation}
where
\begin{equation}
x = s + \sum^{4}_{\mu = 1} y_\mu \hat{\mu}, \quad y_\mu \in \{ 0,1 \}
\end{equation}
are the corners of the hypercube.
Denoting the three indices complementary to $\mu$ by $\alpha$, $\beta$,
$\gamma$ ($\alpha < \beta <\gamma$) we label the corners of the cube
$f(s,\mu) = c(s) \cap c(s-\hat{\mu})$ according to
\begin{equation} \begin{array}{l}
s\hat{=}1, \quad s+\hat{\alpha}\hat{=}2,
\quad s+\hat{\beta}\hat{=}3, \quad
  s+\hat{\alpha}+\hat{\beta}\hat{=}4, \\
s+\hat{\gamma}\hat{=}5, \quad
s+\hat{\alpha}+\hat{\gamma}\hat{=}6, \quad
  s+\hat{\beta}+\hat{\gamma}\hat{=}7, \quad
  s+\hat{\alpha}+\hat{\beta}+\hat{\gamma}\hat{=}8,
\end{array}
\end{equation}
   and take the following interpolation of the section on $f(s,\mu)$
\begin{displaymath} \begin{array}{l} \displaystyle
w^t (s_\mu,s_\alpha,s_\beta,x_\gamma) =
   w^t (s_\mu,s_\alpha,s_\beta,s_\gamma)
   [ w^t (s_\mu,s_\alpha,s_\beta,s_\gamma)^{-1}
   w^t (s_\mu,s_\alpha,s_\beta,s_\gamma+1)
 \\ \displaystyle
\phantom{w^t (s_\mu,s_\alpha,s_\beta,x_\gamma) =} \times
   U^+(1,\gamma) ] ^{y_\gamma} U(1,\gamma)^{y_\gamma} \,, \\
\displaystyle
w^t (s_\mu,s_\alpha+1,s_\beta,x_\gamma) =
   w^t (s_\mu,s_\alpha+1,s_\beta,s_\gamma)
   [ w^t (s_\mu,s_\alpha+1,s_\beta,s_\gamma)^{-1}
 \\ \displaystyle
\phantom{w^t (s_\mu,s_\alpha+1,s_\beta,x_\gamma) =} \times
   w^t (s_\mu,s_\alpha+1,s_\beta,s_\gamma+1)
   U^+(2,\gamma) ] ^{y_\gamma} U(2,\gamma)^{y_\gamma} \,, \\
\displaystyle
w^t (s_\mu,s_\alpha,s_\beta+1,x_\gamma) =
   w^t (s_\mu,s_\alpha,s_\beta+1,s_\gamma)
   [ w^t (s_\mu,s_\alpha,s_\beta+1,s_\gamma)^{-1}
 \\ \displaystyle
\phantom{w^t (s_\mu,s_\alpha,s_\beta+1,x_\gamma) =} \times
   w^t (s_\mu,s_\alpha,s_\beta+1,s_\gamma+1)
   U^+(3 ,\gamma) ] ^{y_\gamma} U(3,\gamma)^{y_\gamma} \,, \\
\displaystyle
w^t (s_\mu,s_\alpha+1,s_\beta+1,x_\gamma) =
   w^t (s_\mu,s_\alpha+1,s_\beta+1,s_\gamma)
   [ w^t (s_\mu,s_\alpha+1,s_\beta+1,s_\gamma)^{-1}
 \\ \displaystyle
\phantom{w^t (s_\mu,s_\alpha+1,s_\beta+1,x_\gamma) =} \times
   w^t (s_\mu,s_\alpha+1,s_\beta+1,s_\gamma+1)
   U^+(4 ,\gamma) ] ^{y_\gamma} U(4,\gamma)^{y_\gamma} \,, \\
\displaystyle
w^t (s_\mu,s_\alpha,x_\beta,x_\gamma) =
   w^t (s_\mu,s_\alpha,s_\beta,x_\gamma)
   [ w^t (s_\mu,s_\alpha,s_\beta,x_\gamma)^{-1}
   w^t (s_\mu,s_\alpha,s_\beta+1,x_\gamma)
 \\ \displaystyle
\phantom{w^t (s_\mu,s_\alpha,x_\beta,x_\gamma) =}  \times
   F^+_{s,\mu}(x_\gamma) ]^{y_\beta}
   F_{s,\mu}(x_\gamma)^{y_\beta} \,,
\end{array}
\end{displaymath}
\begin{equation} \begin{array}{l} \displaystyle
w^t (s_\mu,s_\alpha+1,x_\beta,x_\gamma) =
   w^t (s_\mu,s_\alpha+1,s_\beta,x_\gamma)
   [ w^t (s_\mu,s_\alpha+1,s_\beta,x_\gamma)^{-1}
 \\ \displaystyle
\phantom{w^t (s_\mu,s_\alpha+1,x_\beta,x_\gamma) =} \times
   w^t (s_\mu,s_\alpha+1,s_\beta+1,x_\gamma)
   G^+_{s,\mu}(x_\gamma) ]^{y_\beta}
   G_{s,\mu}(x_\gamma)^{y_\beta} \,, \\
\displaystyle
w^t (s_\mu,x_\alpha,x_\beta,x_\gamma) =
   w^t (s_\mu,s_\alpha,x_\beta,x_\gamma)
   [ w^t (s_\mu,s_\alpha,x_\beta,x_\gamma)^{-1}
   w^t (s_\mu,s_\alpha+1,x_\beta  ,x_\gamma)
 \\ \displaystyle
\phantom{w^t (s_\mu,x_\alpha,x_\beta,x_\gamma) =} \times
   L^+_{s,\mu}(x_\beta,x_\gamma) ]^{y_\alpha}
   L_{s,\mu}(x_\beta,x_\gamma)^{y_\alpha} \,,
\end{array}
\end{equation}
where $y=x-s$ and $t=s$ or $t=s-\hat{\mu}$. Furthermore
\begin{equation} \begin{array}{l} \displaystyle
F_{s,\mu}(x_\gamma) = U^+(1,\gamma)^{y_\gamma}
  [ U(1,\gamma) U(5,\beta) U^+(3,\gamma) U^+(1,\beta)
  ] ^{y_\gamma} U(1,\beta) U(3,\gamma)^{y_\gamma} \,, \\
\displaystyle
G_{s,\mu}(x_\gamma) = U^+(2,\gamma)^{y_\gamma}
  [ U(2,\gamma) U(6,\beta) U^+(4,\gamma) U^+(2,\beta)
  ] ^{y_\gamma} U(2,\beta) U(4,\gamma)^{y_\gamma} \,, \\
\displaystyle
H_{s,\mu}(x_\gamma) = U^+(1,\gamma)^{y_\gamma}
  [ U(1,\gamma) U(5,\alpha) U^+(2,\gamma) U^+(1,\alpha)
  ] ^{y_\gamma} U(1,\alpha) U(2,\gamma)^{y_\gamma} \,,  \\
\displaystyle
K_{s,\mu}(x_\gamma) = U^+(3,\gamma)^{y_\gamma}
  [ U(3,\gamma) U(7,\alpha) U^+(4,\gamma) U^+(3,\alpha)
  ] ^{y_\gamma} U(3,\alpha) U(4,\gamma)^{y_\gamma} \,, \\
\displaystyle
L_{s,\mu}(x_\beta,x_\gamma) = F^+_{s,\mu}(x_\gamma)^{y_\beta}
  [ F_{s,\mu}(x_\gamma) K_{s,\mu}(x_\gamma)
  G^+_{s,\mu}(x_\gamma) H^+_{s,\mu}(x_\gamma) ] ^{y_\beta}
\\ \displaystyle
\phantom{L_{s,\mu}(x_\gamma,x_\beta) =} \times
  H_{s,\mu}(x_\gamma) G_{s,\mu}(x_\gamma)^{y_\beta} \,.
\end{array}
\end{equation}

Now we interpolate the section to the
interior of $c(s)$ as follows:
 \begin{equation}  \begin{array}{l} \displaystyle
 w^s(x_1,x_2,x_3,x_4)  =  w^s(s_1,x_2,x_3,x_4)
 [ w^s(s_1,x_2,x_3,x_4)^{-1}
 w^s(s_1 + 1,x_2,x_3,x_4)
 \\ \displaystyle
\phantom{ w^s(x_1,x_2,x_3,x_4)  = }  \times
  M^{s \,+   }(x_2,x_3,x_4) ] ^{y_1}
  M^s(x_2,x_3,x_4) ^{y_1} \,,
 \end{array}
 \end{equation}
where
 \begin{equation}  \begin{array}{l} \displaystyle
M^s(x_2,x_3,x_4)  =  L^+_{s,1}(x_3,x_4) ^{y_2}
[ L_{s,1}(x_3,x_4)L_{s+\hat{2},2}(x_3,x_4)
L^+_{s+\hat{1},1}(x_3,x_4)
 \\ \displaystyle
\phantom{M^s(x_2,x_3,x_4)  = } \times
L^+_{s,2}(x_3,x_4) ] ^{y_2}
   L_{s,2}(x_3,x_4) L_{s+\hat{1},1}(x_3,x_4) ^{y_2}
 \end{array}
 \end{equation}
is one more interpolating function.

To investigate the properties of the section $w^s$ under gauge
transformations we also interpolate the gauge transformation $g$ from
the corners  of the hypercube to its interior.  The interpolation
of $g$ from the corners of $c(s)$ to its boundary was already given
in ref.~\cite{int}:
\begin{displaymath} \begin{array}{l} \displaystyle
g (s_\mu,s_\alpha,s_\beta,x_\gamma) =
   g (s_\mu,s_\alpha,s_\beta,s_\gamma)
   [ g (s_\mu,s_\alpha,s_\beta,s_\gamma)^{-1}
   g (s_\mu,s_\alpha,s_\beta,s_\gamma+1)
 \\ \displaystyle
\phantom{g (s_\mu,s_\alpha,s_\beta,x_\gamma) =} \times
   U^+(1,\gamma) ] ^{y_\gamma} U(1,\gamma)^{y_\gamma} \,,
\end{array} \end{displaymath}
\begin{equation} \begin{array}{l} \displaystyle
g (s_\mu,s_\alpha+1,s_\beta,x_\gamma) =
   g (s_\mu,s_\alpha+1,s_\beta,s_\gamma)
   [ g(s_\mu,s_\alpha+1,s_\beta,s_\gamma)^{-1}
 \\ \displaystyle
\phantom{g (s_\mu,s_\alpha+1,s_\beta,x_\gamma) =} \times
   g (s_\mu,s_\alpha+1,s_\beta,s_\gamma+1)
   U^+(2,\gamma) ] ^{y_\gamma} U(2,\gamma)^{y_\gamma} \,, \\
\displaystyle
g (s_\mu,s_\alpha,s_\beta+1,x_\gamma) =
   g (s_\mu,s_\alpha,s_\beta+1,s_\gamma)
   [ g (s_\mu,s_\alpha,s_\beta+1,s_\gamma)^{-1}
 \\ \displaystyle
\phantom{g (s_\mu,s_\alpha,s_\beta+1,x_\gamma) =} \times
   g (s_\mu,s_\alpha,s_\beta+1,s_\gamma+1)
   U^+(3 ,\gamma) ] ^{y_\gamma} U(3,\gamma)^{y_\gamma} \,, \\
\displaystyle
g (s_\mu,s_\alpha+1,s_\beta+1,x_\gamma) =
   g (s_\mu,s_\alpha+1,s_\beta+1,s_\gamma)
   [ g (s_\mu,s_\alpha+1,s_\beta+1,s_\gamma)^{-1}
 \\ \displaystyle
\phantom{g (s_\mu,s_\alpha+1,s_\beta+1,x_\gamma) =} \times
   g (s_\mu,s_\alpha+1,s_\beta+1,s_\gamma+1)
   U^+(4 ,\gamma) ] ^{y_\gamma} U(4,\gamma)^{y_\gamma} \,, \\
\displaystyle
g (s_\mu,s_\alpha,x_\beta,x_\gamma) =
   g (s_\mu,s_\alpha,s_\beta,x_\gamma)
   [ g (s_\mu,s_\alpha,s_\beta,x_\gamma)^{-1}
   g (s_\mu,s_\alpha,s_\beta+1,x_\gamma)
 \\ \displaystyle
\phantom{g (s_\mu,s_\alpha,x_\beta,x_\gamma) =}  \times
   F^+_{s,\mu}(x_\gamma) ]^{y_\beta}
   F_{s,\mu}(x_\gamma)^{y_\beta} \,, \\
\displaystyle
g (s_\mu,s_\alpha+1,x_\beta,x_\gamma) =
   g (s_\mu,s_\alpha+1,s_\beta,x_\gamma)
   [ g (s_\mu,s_\alpha+1,s_\beta,x_\gamma)^{-1}
 \\ \displaystyle
\phantom{g (s_\mu,s_\alpha+1,x_\beta,x_\gamma) =} \times
   g (s_\mu,s_\alpha+1,s_\beta+1,x_\gamma)
   G^+_{s,\mu}(x_\gamma) ]^{y_\beta}
   G_{s,\mu}(x_\gamma)^{y_\beta} \,, \\
\displaystyle
g (s_\mu,x_\alpha,x_\beta,x_\gamma) =
   g (s_\mu,s_\alpha,x_\beta,x_\gamma)
   [ g (s_\mu,s_\alpha,x_\beta,x_\gamma)^{-1}
   g (s_\mu,s_\alpha+1,x_\beta  ,x_\gamma)
 \\ \displaystyle
\phantom{g (s_\mu,x_\alpha,x_\beta,x_\gamma) =} \times
   L^+_{s,\mu}(x_\beta,x_\gamma) ]^{y_\alpha}
   L_{s,\mu}(x_\beta,x_\gamma)^{y_\alpha} \,,
\end{array}
\end{equation}
In the interior of $c(s)$ we obtain
 \begin{equation}  \begin{array}{l} \displaystyle
   g(x_1,x_2,x_3,x_4)  =    g(s_1,x_2,x_3,x_4)
 [   g(s_1,x_2,x_3,x_4)^{-1}
   g(s_1 + 1,x_2,x_3,x_4)
\\ \displaystyle
\phantom{g(x_1,x_2,x_3,x_4)  = } \times
  M^{s \,+   }(x_2,x_3,x_4) ] ^{y_1}
  M^s(x_2,x_3,x_4) ^{y_1} \,.
 \end{array}
 \end{equation}
Using the gauge transformation properties of $L_{s,\mu}(x_\beta,
x_\gamma)$,
\begin{equation}
\bar{L}_{s,\mu} (x_\beta,x_\gamma)
= g(s+y_\beta\hat{\beta}+y_\gamma\hat{\gamma})
L_{s,\mu}(x_\beta,x_\gamma) g(s+\hat{\alpha}+y_\beta\hat{\beta}
+y_\gamma\hat{\gamma})^{-1},
\end{equation}
we find
\begin{equation}
\bar{M}^s(x_2,x_3,x_4) = g(s_1,x_2,x_3,x_4) M^s(x_2,x_3,x_4)
g(s_1 + 1,x_2,x_3,x_4)^{-1}.
\end{equation}
Together with the gauge transformation properties of the section at the
boundary of $c(s)$,
\begin{equation} \begin{array}{rcl} \displaystyle
\bar{w}^s(s_1,x_2,x_3,x_4) & = & g(s) w^s(s_1,x_2,x_3,x_4)
g(s_1,x_2,x_3,x_4)^{-1} \,,\\  \displaystyle
\bar{w}^s(s_1 + 1,x_2,x_3,x_4) & = & g(s) w^s(s_1 + 1,x_2,x_3,x_4)
g(s_1 + 1,x_2,x_3,x_4)^{-1}\,,
\end{array}
\end{equation}
this yields the gauge transformation behavior of the section in the
whole hypercube:
\begin{equation}
\bar{w}^s(x_1,x_2,x_3,x_4) = g(s) w^s(x_1,x_2,x_3,x_4)
g(x_1,x_2,x_3,x_4)^{-1}.
\end{equation}

\bibliographystyle{unsrt}

\end{document}